\documentclass[prb,twocolumn,showpacs,superscriptaddress,preprintnumbers,amsmath,amssymb]{revtex4}
\usepackage{dcolumn}
\usepackage{bm}
\usepackage{graphicx}
\usepackage{bm}
\begin{document}
\title{Unusual field dependence of remanent magnetization in granular CrO${_2}$ : possible role of piezomagnetism}
\author{A. Bajpai}
\affiliation{Leibniz-Institute for Solid State and Materials Research, IFW Dresden, 01171 Dresden, Germany}
\author{R. Klingeler}
\affiliation{Leibniz-Institute for Solid State and Materials Research, IFW Dresden, 01171 Dresden, Germany}
\author{N. Wizent}
\affiliation{Leibniz-Institute for Solid State and Materials Research, IFW Dresden, 01171 Dresden, Germany}
\author{A.K. Nigam}
\affiliation{Tata Institute of Fundamental Research, Homi Bhabha Road, Mumbai 400005, India}
\author{S-W. Cheong}
\affiliation{Rutgers Center for Emergent Materials and Department of Physics and Astronomy, Rutgers University, Piscataway, New Jersey, 08854, USA}
\author{B. B\"{u}chner}
\affiliation{Leibniz-Institute for Solid State and Materials Research, IFW Dresden, 01171 Dresden, Germany}
\date{\today}
\begin{abstract}
 
 We present low field thermoremanent magnetization (TRM) measurements in granular CrO${_2}$ and  composites of ferromagnetic (FM) CrO${_2}$ and antiferromagnetic (AFM) Cr${_2}$O${_3}$. TRM in these samples is seen to display two distinct time scales. A quasi static part of remanence, appearing only in low field regime exhibits a peculiar field dependence. TRM is seen to first rise and then fall with increasing cooling fields, eventually vanishing above a critical field.  Similar features in TRM have previously been observed in some antiferromagnets that exhibit the phenomenon of piezomagnetism. Scaling analysis of the TRM data suggest that presumably piezomoments generated in the AFM component drive the FM magnetization dynamics in these granular systems in low field regime.             
\end{abstract}

\maketitle 
For granular specimens of half metallic ferromagnet CrO${_2}$( T${_C}$ $\approx$ 393 K), it has been demonstrated earlier that the growth of an insulating oxide Cr${_2}$O${_3}$  naturally occurs on the surface of the CrO${_2}$ grains\cite{Cheong1, Coey1,Dai1}. This makes granular CrO${_2}$ an attractive magnetoresistive material that exhibits spin polarized tunnelling through the network of metallic grains of CrO${_2}$, separated by insulating grain boundary of Cr${_2}$O${_3}$ \cite{Cheong1, Coey1,Dai1}. This also renders the system a very interesting aspect: a spin polarized metal in contact with a dielectric. It is to be noted that the dielectric under consideration, (i.e. Cr${_2}$O${_3}$ ) in its bulk form is a well known room temperature AFM (T${_N}$ $\approx$ 307 K) and a prototypical magnetoelectric material\cite{Rado, Folen, Martin, Fiebig}. However, the manner in which its AFM character affects the magnetic and transport properties, when it appears as a grain boundary in granular CrO${_2}$, remains largely unexplored. The primary difficulty arises due to the fact that it is hard to track the subtle change in AFM grain boundary in the presence of strongly ferromagnetic CrO${_2}$, at least through routine magnetization measurements. 
      
 A possible means of disentagling the magnetic contribution from FM CrO${_2}$ and AFM Cr${_2}$O${_3}$ is to perform thermoremanent magnetization measurements. Study of remanent magnetization (via cooling a magnetic system through its transition temperature under a fixed magnetic field) reveals important information regarding domain motion, pinning mechanisms and other metastable states.  The efficiency of these measurements in probing antiferromgnets has been demonstrated earlier \cite{Kleemann, Palacio,Mattsson, Fries, Binek}. 

\begin{figure}
\includegraphics[width=7cm,height=6cm]{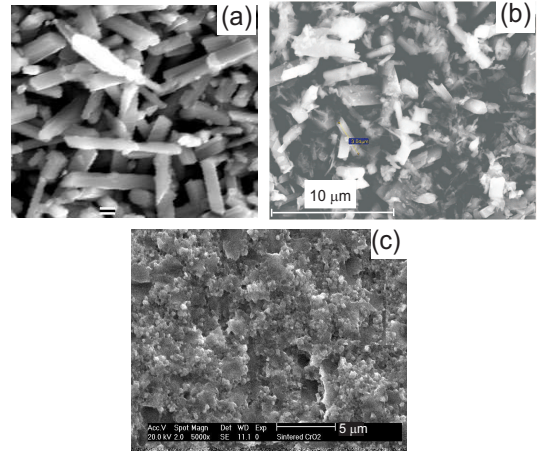}
\caption{ Scanning electron micrographs for CrO${_2}$ samples containing (a) 10 $\%$, (b) 40 $\%$ and (c)  75 $\%$ mass fraction of Cr${_2}$O${_3}$. In (a) and (b), the long micron size rods of CrO${_2}$ grains are visible. On further annealing under high pressure, the grains are seen to coagulate (c). }
\end{figure}	
 
	We investigate granular CrO${_2}$ in which Cr${_2}$O${_3}$ appears as a grain boundary, and highly diluted composites of CrO${_2}$ and Cr${_2}$O${_3}$ with varying mass fraction of Cr${_2}$O${_3}$. These sintered pellets have CrO${_2}$ grains in the form of large micron size rods (5-10 $\mu$m length) \cite{Bajpai2, Bajpai3}.  Figure 1a and Figure 1b are the  scanning electron micrograph  pictures of the sintered pellets, which  contain about 10$\%$  and 40 $\%$ mass fraction of Cr${_2}$O${_3}$ respectively .  On the individual CrO${_2}$ grains (having a well defined rod like  shape), the antiferromagnetic oxide Cr${_2}$O${_3}$ appears as a surface layer (or grain boundary). In this microstructural regime, the samples mimic typical 'metal in insulator scenario' which has also been observed in compacted commercial powders of CrO${_2}$\cite{Coey1,Dai1}. These samples can be termed as 'granular metal' as far as their electron tranport properties are concerned \cite{Coey1, Bajpai1}.  We also investigate a sample which is further sintered under high pressure (6GPa) and high temperature (upto 700 C). The process of high pressure annealing results in a composite with 75 $\%$  mass fraction of Cr${_2}$O${_3}$ and more importantly the  coagulation of  CrO${_2}$ grains (Figure 1c).  Magnetization measurements  on all these samples are done using a Quantum Design SQUID magnetometer. 
	
 In  granular CrO${_2}$ as well as in highly diluted composites, both field and temperature dependence of magnetization (i.e. MH and MT respectively) follow a classical ferromagnetic behaviour arising from CrO${_2}$, thus remaining insensitive to the presence of AFM Cr${_2}$O${_3}$ \cite{Bajpai1}. Figure 2a displays MH isotherm measured at 5K for all the three samples.  In figure 2b,  temperature variation of magnetization measured at various fields for a representative composite with 40 $\%$ Cr${_2}$O${_3}$ is shown. The magnetization is seen to increase with increasing magnetic field. Similar trend is seen in other two samples.  
     
\begin{figure}
\includegraphics[width=8cm,height=4cm]{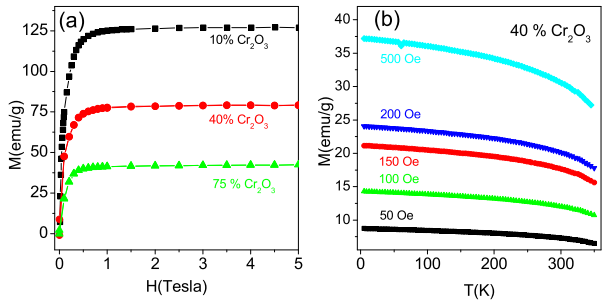}
\caption{(a) MH isotherm recorded at 5K for CrO${_2}$ sample with 10 $\%$, 40 $\%$ and 75 $\%$ mass fraction of Cr${_2}$O${_3}$. (b) Magnetization as a function of temperature in various magnetic fields between 50 to 500 Oe for the sample with 40 $\%$ mass fraction of Cr${_2}$O${_3}$.}
\end{figure}  
  
  In Figure 3, we display TRM measurements for all the three samples. The protocol of the TRM measurements is as follows: the sample is cooled from 350 K to 5K in the presence of a fixed magnetic field, hereby referred to as H$_{cool}$. At 5K, the field is switched off and the remanent magnetization is recorded as a function of temperature in heating cycles.  We emphasize that as soon as the magnetic field is switched off, the magnetization is seen to decay almost instantly with time. However, it eventually arrives at some residual value where it stays without a notable decay. For instance, we take note from figure 2b, the 'infield' magnetization value for the 40 $\%$ composite, which is around 15 emu/g at 5 K at measuring field of 100 Oe. On switching off the field at 5K, the magnetization attains a residual value of about 5 emu/g (Figure 3b). This 'residual' remanent magnetization is fairly long lived in experimental time scale. What we have plotted in figure 3 is the temperature variation of this quasi static remanence, measured for various  H$_{cool}$. 
   
Figure 3a displays temperature variation of TRM  on the sample with 10 $\%$ mass fraction of Cr${_2}$O${_3}$, for a few selected values of H$_{cool}$. The magnitude of TRM first increases with increasing Hcool and attains its maximum value at a critical cooling field. Above this field, the TRM is seen to decrease in magnitude with increasing H$_{cool}$ and eventually vanishes at cooling fields of about 1 kOe. For all our measurements, the residual field of SQUID magnetometer (which is estimated to be of the order of 2-5 Oe during the measurement) is not seen to interfere with this unusual field dependence of TRM.  For the same sample, we also conducted TRM measurements in the high temperature oven in squid magnetometer so as to heat the sample above 400 K (T${_C}$ of the FM grain) after each measurement. This data set is displayed in the inset of Figure 3a.  Here the H$_{cool}$ was applied at 450 K which is much above the T${_C}$ of FM grain and the T${_N}$ of the AFM grain boundary. Subsequently, the sample is cooled up to 300 K and the TRM is measured as described earlier.  The TRM and its unusual field dependence is seen to persist right up to 400 K, the T${_C}$ of FM CrO${_2}$.  Qualitatively similar features are observed when the direction of H$_{cool}$ is reversed. 

\begin{figure}
\includegraphics[width=6cm,height=15cm]{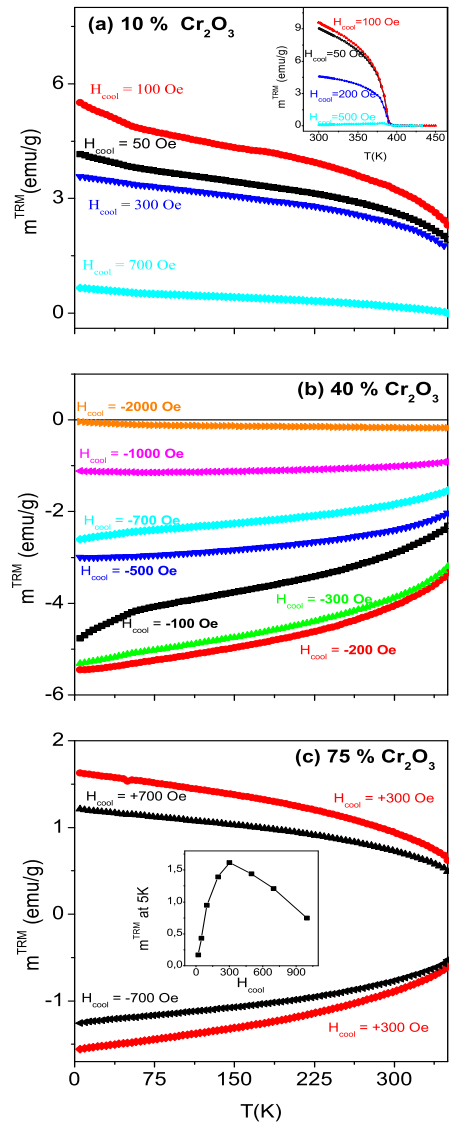}
\caption{: (a) TRM as a function of temperature for  CrO${_2}$ samples containing (a) 10 $\%$ (b) 40 $\%$ and (c) 75 $\%$ mass fraction of  Cr${_2}$O${_3}$ in various H$_{cool}$.  The inset in (a)  shows the same in the temperatures between 300 K to 450 K for the 10 $\%$ sample.  Inset in (c) shows TRM values at 5K plotted as a function of H$_{cool}$ for the sample with 75 $\%$ mass fraction of Cr${_2}$O${_3}$.}
\end{figure}   
   
   These experimental signatures are reproduced  in  composites of CrO${_2}$/Cr${_2}$O${_3}$,  irrespective of the mass fraction of Cr${_2}$O${_3}$ or the micostructure, as is evident from Figure 3b and Figure 3c. Figure 3b displays the TRM measured in negative cooling fields for a composite containing 40 $\%$ mass fraction of Cr${_2}$O${_3}$. The grain size of CrO${_2}$  in this sample was similar to the sample with 10 $\%$ mass fraction of Cr${_2}$O${_3}$. Figure 3c displays the symmetric nature of TRM in positive and negative cooling fields for the 75 $\%$ composite.  The unusual field dependence of TRM is clearly depicted in the inset of Figure 3c where the TRM at 5K is plotted against the H$_{cool}$ from the cross section of various temperature scans for the 75 $\%$ composite.  We emphasize that the unusual field dependence of TRM (inset of figure 3c) is qualitatively similar to what is seen in other two samples. However the field at which the TRM reaches its maximum and also the field at which it vanishes, varies slightly from sample to sample. The data obtained on a variety of samples indicate that the peculiar field dependence of TRM is intrinsic to these materials. More importantly, it appears only in the low field limit.     
   
	The unusual field dependence of TRM is not reflected in the routine 'in field' magnetization in any of these samples. In all these samples, the magnetization increases in magnitude with increasing applied field (Figure 2b) and saturates at about 1 Tesla (Figure 2a), consistent with FM CrO${_2}$ \cite{Bajpai1}. Thus it is clear that while regular M vs T data exhibit no surprises, for the similar measurement protocol, a part of TRM exhibits an unusual field dependence.  This  field dependence of TRM is strikingly different from what is usually seen in TRM arising in a conventional ferromagnet. It is clear that a part of the remanence, which exhibits instant decay after switching off the field, arises from the regular domain wall dynamics intrinsic to FM CrO${_2}$. However, the quasi static remanance appears to be intimately connected to the magnetization at the AFM/FM interface. 
    
    A possible scenario is that the AFM phase pins the FM spins at the interface. However this pinning does not seem to appear from exchange bias alone. Though, exchange bias is certainly expected at the FM/AFM interface but this contribution to TRM should have vanished around 300 K (near the T${_N}$ of Cr${_2}$O${_3}$). Moreover, the exchange bias field is inversely proportional to the magnetization of the FM layer\cite{Mieklejohn}. It is evident that within this simplistic picture, the peak in TRM at moderate fields of the order of 100 -200 Oe and also vanishing of TRM at fields of the order of 1 -2 kOe can not be understood. Thus, though exchange bias effects at the FM/AFM interface may be present, the observed TRM and the related pinning mechanism does not seem to arise from it.  We also recall that a spin flop phase in Cr${_2}$O${_3}$ is known to occur at fields of the order of  5.8 Tesla and temperatures below 90K \cite{Fiebig}. However, the unusual field dependence of TRM (in a wide temperature range between 5K-400K) can not be understood by invoking spin flop in the AFM layer.  Besides the fact that a field induced spin flop transition is extremely unlikely at such moderate fields, neither the field dependence nor the vanishing of TRM can be explained by this scenario.
    
    It is important to recall that similar features in TRM, especially in low field regime, have previously been observed in certain antiferromagnets whose origin could not be understood within the conventional relaxation processes occurring either in long range order or in the metastable systems\cite{Kleemann, Palacio,Mattsson, Fries, Binek}. This remanence, has been experimentally observed in Fe$_{0.47}$Zn0$_{0.53}$F${_2}$ \cite{Kleemann}, Mn$_{1-x}$Zn${_x}$F${_2}$, K${_2}$Fe$_{1-x}$In${_x}$Cl${_5}$ etc.\cite{Fries}  and more recently in the epitaxial thin films of Cr${_2}$O${_3}$ \cite{Binek}.  In case of FeZnF$_{2}$ \cite{Kleemann} as well as pure  Cr${_2}$O${_3}$ films \cite{Binek}, this peculiar remanance  has been associated with piezomagnetically frozen moments which are, from symmetry arguments, allowed in these antiferromagnets. Here, piezomagnetism is defined as a third rank tensor (P$_{ijk}$) relating the component of magnetization (M$_{i}$) induced in 'i' direction due to the applied stress $\sigma$$_{jk}$ (M$_{i}$ = P$_{ijk}$ $\sigma$$_{jk}$) \cite{Phillips}.  The stress can be applied externally or it can also be intrinsic in the system\cite{Kleemann}. In case of epitaxial thin films of Cr${_2}$O${_3}$ stress is understood to arise from the lattice mismatch of the film with the substrate\cite{Binek}.  We also emphasize, that in context of spintronic devices based on multiferroic materials which exhibit electric field driven magnetization and magnetic field driven polarization, the phenomenon of piezomagnetism adds additional possibility of strain/stress mediated magnetoelectric coupling \cite{Ramesh}.
    
    In  our CrO${_2}$/Cr${_2}$O${_3}$ samples, we note that the magnitude of TRM obtained is large and it vanishes only above 400 K which is the actual FM T${_C}$, thus indicating that it primarily arises from the ferromagnetic grains. However qualitative features of TRM, particularly its field dependence is remarkably similar to the TRM data obtained on thin films of AFM Cr${_2}$O${_3}$\cite{Binek}.  This correlation strongly suggest that at least low field TRM in our samples is modulated by the grain boundary and gives rise to an unconventional pinning mechanism at the FM/AFM interface.  It appears that in low field regime, this pinning may arise exclusively due to the piezomoments generated in the AFM Cr${_2}$O${_3}$. These piezomoments can arise due to the natural stress at the interface coming from the lattice mismatch between CrO${_2}$ and Cr${_2}$O${_3}$. This stress can induce a finite piezomoment in the AFM layer as this phenomenon is consistent with the crystal symmetry of Cr${_2}$O${_3}$.  In case of thin films of Cr${_2}$O${_3}$, the process of field cooling creates AFM domain whose direction is determined by the direction of local stress and the Zeeman energy considerations \cite{Binek}. As in case of thin films of pure Cr${_2}$O${_3}$, these effects appear to peak and saturate in moderate field range of a few 100 Oe.  It has been shown in films of Cr${_2}$O${_3}$, that the pinned moment follows similar field dependence (first increasing with increasing field and then saturating at a few 100 Oe) even above the T${_N}$ though  the magnitude of this effect is smaller\cite{Binek}. The mechanical stress arising due to the lattice mismatch is likely to be finite above T${_N}$.  It is also to be recalled that in case of pure Cr${_2}$O${_3}$, the atomic ME coefficient may exist in the paramagnetic regime but the net effect enhances when long range AFM order sets in \cite{Martin}. 
    
	The generation of finite piezomoment in the epitaxial thin films of pure Cr${_2}$O${_3}$ has been concluded from both the functional form, as well as the observation of a scaling behaviour in TRM. Such scaling is understood to arise from the factorization of  the piezomoment, due to which magnetization  is a product function of field and temperature according to the equation m(H,T) = f(H)g(T) where f and g are functions of field and temperature \cite{Binek}. Scaling is achieved by normalization of moment to its value at any fixed temperature which is less than T$_{N}$. Interestingly, we also find that TRM data recorded at various H$_{cool}$, when normalized by TRM at any temperature above 200K, falls into a single curve.  Figure 4 displays TRM normalized to its value measured at 300K for sample with 10 $\%$ mass fraction of Cr${_2}$O${_3}$. The inset shows the scaling for the TRM measured between 5 K-350 K for the same sample.  Here the normalization is done with TRM value measured at 240 K. Deviations from the scaling are observed for H$_{cool}$ more than 200-300 Oe or when TRM is normalized with its value below 200K. Observation of the scaling in this field and temperature window also indicates that possibly similar piezomagnetic traits of AFM Cr${_2}$O${_3}$ leads to the unusual magnetization dynamics in granular CrO${_2}$.
	
	\begin{figure}
\includegraphics[width=6cm, height=5cm]{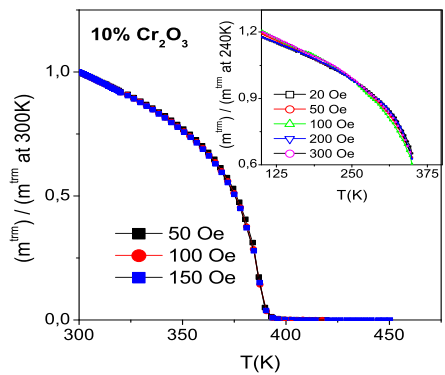}
\caption{Main panel shows normalized TRM in the temperature range between 300 to 450 K exhibiting a scaling behaviour  in various cooling fields, H$_{cool}$. The inset shows the normalized TRM for data set recorded between 5K to 350 K.  It is seen that for lower H$_{cool}$, when normalized with any temperature above 200 K, the data collapses into a single curve.} 
\end{figure}    

 Once pinned from the process of field cooling through the T${_N}$ (307 K) of Cr${_2}$O${_3}$, the TRM is retained right up to 400 K, and also preserves its peculiar field dependence much above the T${_N}$ of Cr${_2}$O${_3}$. This feature can have interesting implications due to the fact that the magnitude of the pinned moment is significantly large, unlike what is seen in epitaxial thin films of Cr${_2}$O${_3}$. 
 
	In conclusion we present low field thermoremanent magnetization data on the composites of CrO${_2}$/Cr${_2}$O${_3}$. This measurement spectacularly brings out that the effects of magnetization of AFM part, which is not visible in routine 'infield' magnetization in granular CrO${_2}$ sample. The data also indicates that presumably the stress effects, which can naturally occur at the interface of this spin polarized metal (CrO${_2}$) and magnetoelectric insulator (Cr${_2}$O${_3}$) due to the lattice mismatch, may give rise to a finite piezomoment in Cr${_2}$O${_3}$. This, in the low field limit modulates the magnetization of CrO${_2}$. Once frozen from the process of field cooling, these piezomoment can pin the magnetization of FM spins at the interface and thus open up additional possibilities of tuning the spintronic devices via stress mediated coupling in granular CrO${_2}$. 

	 A.B. acknowledges support through the EU Marie Curie IIF fellowship under project 040127-NEWMATCR. Work at Rutgers was supported by NSF-DMR-0520471.

\end{document}